\documentclass{pasj00}

\SetRunningHead{Richmond et al.}{Proper motions with Subaru}

\usepackage{times}

\begin{document}

\title{Proper motions with Subaru I. Methods 
           and a first sample in the Subaru Deep Field}
\author{Michael W. \textsc{Richmond}%
}
\affil{%
   Physics Department, Rochester Institute of Technology, \\
   Rochester, NY 14623, USA}
\email{mwrsps@rit.edu}
\author{Tomoki \textsc{Morokuma}}
\affil{Optical and Infrared Astronomy Division, National Astronomical Observatory of Japan, \\
        2-21-1 Osawa, Mitaka, Tokyo 181-8588, Japan}
\email{tomoki.morokuma@nao.ac.jp}
\author{Mamoru \textsc{Doi}}
\affil{Institute of Astronomy, Graduate School of Science, University of Tokyo, \\
          2-21-1, Osawa, Mitaka, Tokyo 181-0015, Japan}
\email{doi@ioa.s.u-tokyo.ac.jp}
\author{Yutaka \textsc{Komiyama}}
\affil{Optical and Infrared Astronomy Division, National Astronomical Observatory of Japan, \\
        2-21-1 Osawa, Mitaka, Tokyo 181-8588, Japan}
\email{komiyama@subaru.naoj.org}
\author{Naoki \textsc{Yasuda}}
\affil{Institute for Cosmic Ray Research, University of Tokyo, \\
             5-1-5 Kashiwa-no-Ha, Kashiwa, Chiba 277-8582, Japan}
\email{yasuda@icrr.u-tokyo.ac.jp}
\author{Sadanori \textsc{Okamura}}
\affil{Department of Astronomy and Research Center for the Early Universe,
            School of Science, University of Tokyo \\
        7-3-1 Hongo, Bunkyo, Tokyo 113-0033, Japan}
\email{okamura@astron.s.u-tokyo.ac.jp}
\and
\author{Avishay Gal-Yam}
\affil{Benoziyo Center for Astrophysics, Weizmann Institute of Science,
76100 Rehovot, Israel}
\email{avishay.gal-yam@weizmann.ac.il}

\KeyWords{stars: kinematics ${}$ --- Galaxy: kinematics and dynamics${}$ --- Galaxy: structure${}$}

\maketitle

\begin{abstract}
We search for stars with proper motions in a set of twenty 
deep Subaru images, covering about 0.28 square degrees
to a depth of $i' \simeq 25$, 
taken over a span of six years.
In this paper, we describe in detail our reduction 
and techniques to identify moving objects.
We present a first sample of
99 stars with motions of high significance,
and discuss briefly the populations
from which they are likely drawn.
Based on photometry and motions alone,
we expect that 9 of the candidates
may be white dwarfs.
We also find a group of stars which may 
be extremely metal-poor subdwarfs in the halo.
\end{abstract}

\section{Introduction}

The basic structure of our Milky Way galaxy seems clear:
a thin disk of young stars, gas and dust circles the
center quietly, immersed within a thicker disk of older stars.
Both disks sit inside a nearly spherical halo
of very old, metal-poor stars which do not share the
overall rotation of the disks.  
Surrounding everything is an extended distribution
of dark matter.
Our knowledge of the details within this big picture,
on the other hand, is not so clear.
Recent large-scale projects, such as the 
Sloan Digital Sky Survey,
have measured the properties of high-luminosity
stars throughout the halo
(see, for example, \cite{Yanny2000}, 
and \cite{Juric2008}),
while the rapid development of infrared 
detectors has allowed projects such 
as 
the Two Micron All Sky Survey
(\cite{Skrutskie2006})
and Spitzer Space Telescope
(\cite{Patel2004})
to pierce the dusty disks and 
measure the properties of their stars.
Nonetheless,
some portions of the Milky Way
remain largely unexplored.

In particular,
we know little of 
the low-mass, low-luminosity stars
of the halo.
The white dwarfs and metal-poor
subdwarfs of the halo glow so faintly,
from so great a distance,
that they are rarely seen
and more rarely recognized.
As a recent review
(\cite{Reid2005})
points out,
however,
these shy and elusive stars
may dominate the microlensing
events observed towards the
Galactic bulge and the Magellanic Clouds.

There are two ways to search for 
these stars:
cover a very large area on the sky 
to a shallow depth,
or use a ``pencil-beam'' survey to examine
a tiny region much more deeply.
The first approach
(see, for example,
\cite{Oppenheimer2001}
and
\cite{Carollo2006})
will find objects in many directions,
but only out to a small distance from the Sun;
the second approach
(see, for example,
\cite{Mendez2002},  
\cite{Nelson2002},  
and
\cite{Kalirai2004})
probes farther into the halo, but only
in a specific direction.
One way to characterize surveys is to 
combine their area 
with the distance out to which 
they would detect 
some specific star 
to generate
an ``effective volume'' for that type
of star.
In Table
\ref{table:survey_volume},
we compare the projects mentioned
above by this metric, using 
a star of absolute magnitude $M_V = 16.5$,
appropriate for a cool white dwarf.
We assumed a color 
$(V-R) = 0.5$ to convert limiting magnitudes
for $R$-based surveys to $V$-band.

\begin{longtable}{l l l r}
  \caption{Effective volumes, for $M_V = +16.5$}
  \label{table:survey_volume}
  \hline \hline
  Survey & Area (sq.deg.) & limiting mag $V$ & volume (pc$^3$) \\
  \endfirsthead
  \hline \hline
  \endhead
  \hline 
  \endfoot
  \hline 
  \endlastfoot

  \hline 
  Oppenheimer et al. & 4165                & \quad 19.8     & 80000 \\
  Carollo et al.     & 1150                & \quad 20       & 15000 \\
  Nelson  et al.     & \phantom{416}0.021  & \quad 26.5     &  2100 \\
  Mendez             & \phantom{416}0.0013 & \quad 26.0     &    64 \\
  Kalirai et al.     & \phantom{416}0.0031 & \quad 29       &  9800 \\
  this work          & \phantom{416}0.28   & \quad 26       & 14000 \\
  
\end{longtable}

Our project could be described as
a ``pencil-beam'' survey, 
but it uses a very thick pencil.
We examine
images in the area of the Subaru Deep Field (SDF)
(\cite{Kashikawa2004})
acquired over a period of six years
to search for moving objects.
These images were acquired primarily to study
high-redshift galaxies
(\cite{Nagao2004}, \cite{Shimasaku2006}),
but have also been used to find
high-redshift supernovae
(\cite{Poznanski2007}).
The images are nearly as deep as those in some
HST-based surveys, but cover a significantly wider area,
yielding a large effective volume.
We can refer to the SDF catalogs compiled
by 
\citet{Kashikawa2004}
and
\citet{Richmond2005}
for information on our candidates
in multiple passbands ($B$, $V$, $R_c$, $i'$ and $z'$).
Unlike most other pencil-beam surveys,
we have measurements at many epochs:
our dataset contains images taken on 20 nights.
We can therefore measure the proper motion
of our candidates very well,
and place strong constraints on the 
uncertainties in our measurements.

This paper is the first in a series 
on proper motions in several small fields
studied with Subaru.
We will concentrate on techniques,
leaving detailed analysis of the results
for later papers.
In Section 2,
we describe the observations,
their reductions,
and the combination of individual frames
into a single combined image for each night.
In Section 3,
we walk through our procedure for
finding moving objects,
and discuss our criteria for separating
good candidates from bogus ones;
we end up with a first sample of stars
which have very well measured proper motions.
In Section 4, 
we compute the reduced proper motions for 
objects in this sample, and compare their properties
to those of objects drawn from a simulated 
survey of our field.
Finally, in Section 5,
we list our plans for future work on
this dataset, and in other fields with
multiple epochs of deep Subaru imaging.

Astronomers at the Observatoire de Besan\c{c}on 
have created a model of the stellar populations 
in the Milky Way
(\cite{Robin2003})
with a very 
convenient web-based interface\footnote{http://bison.obs-besancon.fr/modele}.
The model consists of four populations of stars --
thin disk, thick disk, spheroid, outer bulge
(the innermost portions of the Milky Way are poorly constrained) --
plus white dwarfs added to each component separately.
The parameters of each component are adjusted 
to produce the best fit to the observed stellar 
populations and their dynamics.
Recent work by 
\citet{Ibata2007}
finds that 
the Besan\c{c}on model
does a very good job of reproducing
observed star counts in two of three deep
fields down to $i_{0} = 24$.
Those authors
criticized
the Besan\c{c}on populations
for being too sharply defined,
forcing them to smooth the model
colors by small amounts ($\sim 0.10$ mag)
in order to match observed color-magnitude
diagrams.
However, since our main concern is to classify
objects very broadly 
using a mixture of kinematics, magnitudes and colors,
we will adopt 
the Besan\c{c}on model
and use it as a reference
throughout this paper
to help us interpret properties of our sample.

\section{Observations}

The SDF is a region at high galactic latitude
($l = 37^{\circ}_{.}6, b = +82^{\circ}_{.}6$)
roughly half a degree on a side.
\citet{Kashikawa2004}
describe very deep optical images 
taken with the Subaru 8.2-meter telescope
and 
Suprime-Cam camera
(\cite{Miyazaki2002}).
We investigated this region using a set of
$i'$-band images with shorter exposure times.
Table
\ref{table:journal}
lists the date for all nights used in our analysis.
Note that our images taken on 2003 April 30 had a
shorter exposure time than the rest; since these images
also had some of the worst seeing, we gave those
measurements very little weight in the final proper motion
calculations.
We split the images taken on the night of 2006 May 3 
into two sets and treated each independently, 
as if taken on different nights.
Since the data taken on 2007 Feb 15 had the best seeing
and the largest number of detected objects,
we adopted it as the fiducial set 
for matching
(see Section 3).

\begin{longtable}{l c r}
  \caption{Observations of the SDF in $i'$-band}
  \label{table:journal}
  \hline \hline
  UT Date \quad  & Julian Date - 2,450,000 & 
             \phantom{} Exptime (seconds) \phantom{} \\
  \endfirsthead
  \hline \hline
  \endhead
  \hline 
  \endfoot
  \hline 
  \endlastfoot

  \hline 
  2001 April 25  & 2024.34  &  3600 \qquad\qquad  \\
  2001 May 20    & 2049.44  &  3240 \qquad\qquad \\ 
  2002 April 12  & 2376.44  &  9780 \qquad\qquad \\
  2002 May 7     & 2401.52  &  5670 \qquad\qquad \\
  2003 April 1   & 2730.43  & 12330 \qquad\qquad \\
  2003 April 3   & 2732.29  &  2730 \qquad\qquad \\
  2003 April 25  & 2754.42  &  4800 \qquad\qquad \\
  2003 April 26  & 2755.36  &  4680 \qquad\qquad \\
  2003 April 30  & 2759.60  &   964 \qquad\qquad \\
  2003 May 1     & 2760.39  &  4583 \qquad\qquad \\
  2005 March 5   & 3434.54  &  5100 \qquad\qquad \\
  2005 March 6   & 3435.54  &  5400 \qquad\qquad \\
  2006 May 3(a)  & 3858.41  &  3000 \qquad\qquad \\
  2006 May 3(b)  & 3858.52  &  2400 \qquad\qquad \\
  2007 February 13  & 4144.59 &   2700 \qquad\qquad \\
  2007 February 14  & 4145.59 &   4200 \qquad\qquad \\
  2007 February 15  & 4146.59 &   4500 \qquad\qquad \\
  2007 February 16  & 4147.60 &   3600 \qquad\qquad \\
  2007 May 16       & 4236.32 &   4180 \qquad\qquad \\
  2007 May 17       & 4237.27 &   3780 \qquad\qquad \\
  
\end{longtable}


During each night of observing, we took a series
of short (typically 180-second to 360-second) exposures,
shifting the telescope position slightly to fill in
small gaps between the ten CCDs on the focal plane.
Using the 
{\it SDFRED} package
(\cite{Ouchi2004})
and NEKO software
(\cite{Yagi2002}),
we followed the procedures described in
section 4 of 
\citet{Kashikawa2004}
to turn all the raw frames taken during the night
into a single, large mosaic.
Briefly, we cleaned the raw images by
subtracting a bias deduced from the overscan regions
and 
dividing by a normalized flatfield frame 
made from a median of many night-time target images.
Using the parameters derived in 
\citet{Miyazaki2002},
we corrected for optical distortions in the focal plane.
Images from all chips were convolved to form
a uniform point-spread function (PSF) across the entire array.
We determined a sky background by calculating 
the local sky at a series of grid points spaced at
intervals of roughly $51$ arcseconds
and using bi-linear interpolation between the grid points;
we then subtracted this sky background from each image.
We used stars shared by adjacent CCDs to determine
the weights to use when combining data from individual
images to make the final mosaic.
The result for each night 
is one 
(or, in the case of 2006 May 3, two) 
large image covering the entire SDF.

The quality of final combined images
varied from night to night.
The Full Width at Half Maximum (FWHM)
ranged from 
$0{\rlap.}^{''}75$ 
to 
$1{\rlap.}^{''}30$,
but, since the plate scale was 
$0{\rlap.}^{''}202$
per pixel, no data was undersampled.
The limiting magnitude also varied with
the conditions,
but was usually 
$i' \sim 25.5$.

\section{Searching for candidates with proper motion}

Selecting objects with proper motions 
from a set of images requires several steps:
finding and measuring the properties of stars in
individual images,
matching stars found at different epochs,
computing the change in position of each star over time,
and deciding which changes are due to real movement.
We will now describe these steps in detail.

In order to find star-like objects in each image,
we used the 
``stars''
program within the 
{\it XVista} package
(\cite{Treffers1989})\footnote{http://spiff.rit.edu/tass/xvista}.
The position of each object was
calculated by fitting a gaussian to the 
background-subtracted, intensity-weighted marginal
sums in each direction
(see \citet{Stone1989} for details).

It is not crucial to separate stars from galaxies
at this early stage, since we will later discard any
objects which do not move significantly;
therefore, we accepted any object with a sharp core,
$0{\rlap.}^{''}6 < {\rm FWHM\ } < 1{\rlap.}^{''}4$,
as a ``star.''
The number of ``stars'' found each night 
ranged from about 20,000 to about 100,000,
depending on the exposure time and seeing.

The Suprime-Cam field is wide enough
that even small uncorrected distortions 
near the edge of the field
might move the apparent position of a star 
enough from one epoch to the next to hide
real, but small, proper motions.
In order to reduce any residual distortions,
we broke the field into smaller units
we will call ``sectors.''
Each sector is a square $1000 \times 1000$ pixels,
or $202 \times 202$ arcseconds, 
on a side.
We allowed a small overlap of 
$10$ arcseconds between adjacent sectors 
so that stars near the edges would not be missed.

\begin{figure}
  \begin{center}
    \FigureFile(80mm,80mm){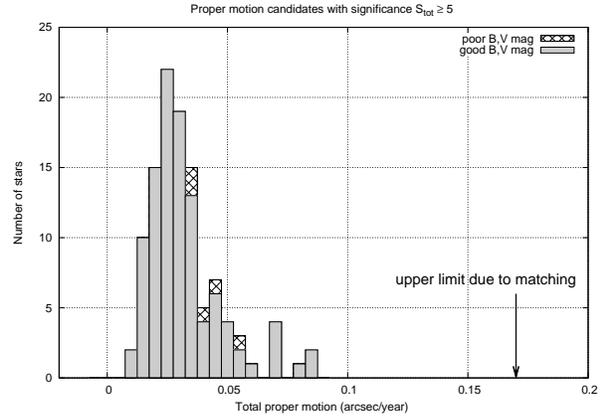}
  \end{center}
  \caption{Distribution of proper motions for
             candidates with high significance.
             \label{fig:pm_histogram} }
\end{figure}

We designated one epoch,
2007 February 15, 
as ``fiducial,''
to serve as the basis of our matching procedure.
For every other image,
we used the
{\it match} package
(\cite{Droege2006})\footnote{http://spiff.rit.edu/match}
to match the objects in each sector
to objects in the corresponding sector of
the fiducial image.
In order to count as an initial match to the fiducial
image,
a star had to lie within 
$1{\rlap.}^{''}0$
of the position of an object in the fiducial frame;
we imposed this limit in order to avoid spurious
matches between unrelated objects.
Given the six-year span of our survey,
this places an upper limit of about
$0{\rlap.}^{''}17$
per year on our proper motion candidates.
We may increase this limit to look for
fast-moving objects in the future.
We will demonstrate later
(see Figure \ref{fig:pm_histogram})
that this requirement does not have a strong
effect on the results.

There were typically three hundred to eight hundred 
pairs of matching items found within each sector.
We transformed the (pixel) coordinates of each star
to the (pixel) coordinates of the fiducial image
in the following iterative manner.
First, we used all the matched pairs
in the sector to find the coefficients
of a linear transformation
\begin{eqnarray}
x' &=  A + Bx + Cy \\
y' &=  D + Ex + Fy 
\end{eqnarray}
via a least-squares technique.
Next, we computed the residuals between the
positions of the members of each pair
in the fiducial coordinate system.
We discarded pairs with large residuals;
specifically, any pair with a residual
more than 10 times the 35th percentile.
We then went back to compute
new coefficients of the linear transformation
with the surviving pairs.
We repeated this procedure three times
in each sector.
Ignoring a few sectors with very few objects, 
the mean residual difference in position 
for surviving items matched to the fiducial frame
was
$0{\rlap.}^{''}071$.
However, most of the objects contributing to this
residual, like most of the objects in each image,
are faint, 
and some of the matches are spurious.
The uncertainties in the positions of bright 
objects are considerably smaller,
as we will show below.

We performed trials using a cubic transformation
between the two coordinate systems,
but found that the residuals were not significantly
smaller than those based on a linear transformation.

The final steps in our matching procedure
were to discard duplicate entries for objects in the
overlapping areas between sectors,
and to discard any objects which appeared in
fewer than five epochs.
The result was a set of positions 
in the fiducial coordinate system
for objects appearing in at least five epochs.
In order to estimate the uncertainty
in the calculated positions,
we computed the mean position of each objects
in each coordinate (row and col)
and its standard deviation;
we then discarded measurements more than two standard 
deviations from the mean and recalculated mean and standard deviation.
As shown in the first two rows of Table \ref{table:pos_uncert},
the typical clipped standard deviation rose from
$0{\rlap.}^{''}007$ for bright, unsaturated objects
to 
$0{\rlap.}^{''}047$ for faint objects.

\begin{longtable}{l c r r r r r r r}
  \caption{Estimates of uncertainty in position (arcsec) as function of $i'$-band mag}
  \label{table:pos_uncert}
  \hline \hline
  Sample, method  & direction & $19-20$ & $20-21$ & $21-22$ & $22-23$ & 
                           $23-24$ & $24-25$ & $25-26$ \\
  \endfirsthead
  \hline \hline
  \endhead
  \hline 
  \endfoot
  \hline 
  \endlastfoot

  \hline 
  All objects, clipped & row & 0.010  & 0.006 & 0.006 & 0.007 & 0.012 &
                    0.028 & 0.047 \\
  stdev from mean pos & col & 0.011  & 0.007 & 0.007 & 0.008 & 0.013 &
                    0.028 & 0.047 \\

   \quad & & & & & & & &  \\

  Moving candidates,  & row & 0.007  & 0.007 & 0.007 & 0.009 & 0.022 &
                    0.036 & 0.052 \\
  scatter from fit    & col & 0.008  & 0.008 & 0.009 & 0.011 & 0.023 &
                    0.036 & 0.052 

\end{longtable}

\begin{figure}
  \begin{center}
    \FigureFile(80mm,80mm){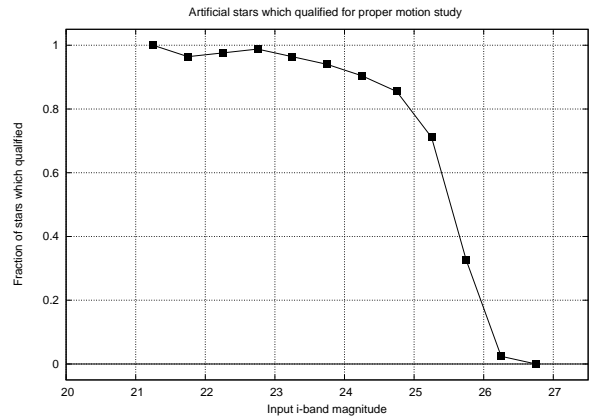}
  \end{center}
  \caption{Fraction of artificial stars added to the images
             which were detected and matched in at least
             5 epochs.  \label{fig:completeness} }
\end{figure}

In order to create a sample of objects for which 
proper motions could be measured accurately,
we selected all objects which appeared in 
the fiducial image and at least four others.
A total of $79605$ objects satisfied
this requirement.
Faint objects were less likely to be selected,
since they might not be detected on nights
with poor seeing.
In order to check the completeness of 
this sample as a function of magnitude,
we inserted a set of $1000$ artificial stars
with magnitudes ranging from
$21 < i' < 27$
into the images.
We then re-analyzed the entire set of images
as before.
Figure \ref{fig:completeness}
shows the fraction of artificial stars
which were detected and placed into
the sample for further study.
Since the fraction falls to 50\%
at 
$i' \simeq 25.5$,
we estimate that our search may be considered
complete to that magnitude.

We subjected this sample to a 
round of tests.
For each coordinate, row and column,
we made a linear fit to position as a function
of Julian Date.
Our fitting routine,
following 
\citet{Press1992},
provides values for the 
the slope
$b$
of this line,
the 95 percent confidence interval
$ci$
in the value of the slope,
and the scatter 
$s_x$ around the line.
The scatter is another estimate
of the one-dimensional uncertainty in
the position of a single measurement;
we show its values in
the lower rows of Table \ref{table:pos_uncert}.

In order to verify that our fitting method yields both
the correct motion and an appropriate uncertainty,
we ran a Monte Carlo simulation.
For each integer magnitude between $21 < i' < 25$,
and for each value of 1-D annual proper motion 
$\mu = 0{\rlap.}^{''}01, 0{\rlap.}^{''}02, \ldots,  0{\rlap.}^{''}20$,
we created 100 artificial stars.
For each star,
we drew 15 epochs randomly from our list of observations 
(see Table \ref{table:journal}),
and computed a set of positions,
using the true proper motion plus some random error
in each direction drawn from a gaussian distribution
consistent with our measurements of $s_x$ for the given magnitude.
We then submitted this list of simulated positions
to our fitting routines,
and compared their results to the true proper motions.
We found that over this entire range of magnitudes
and motions,
our estimates for the proper motion
and its uncertainty
were accurate.

Next, we computed a significance of the slope
for each coordinate:

\begin{eqnarray}
S_{row} &\equiv { {b_{row}} \over {{ci}_{row} } } \\
S_{col\phantom{}} &\equiv { {b_{col}} \over {{ci}_{col} } }
\end{eqnarray}
Choosing objects based on the significance
of their motion in one direction alone would
discriminate against objects moving
diagonally across the CCD,
which was aligned with the equatorial coordinate system.
Therefore, we combined the significance values 
to create an unbiased measure of motion,
$S_{tot} = \sqrt{ {S_{row}^2 + S_{col}^2 } }$.

\begin{figure}
  \begin{center}
    \FigureFile(110mm,110mm){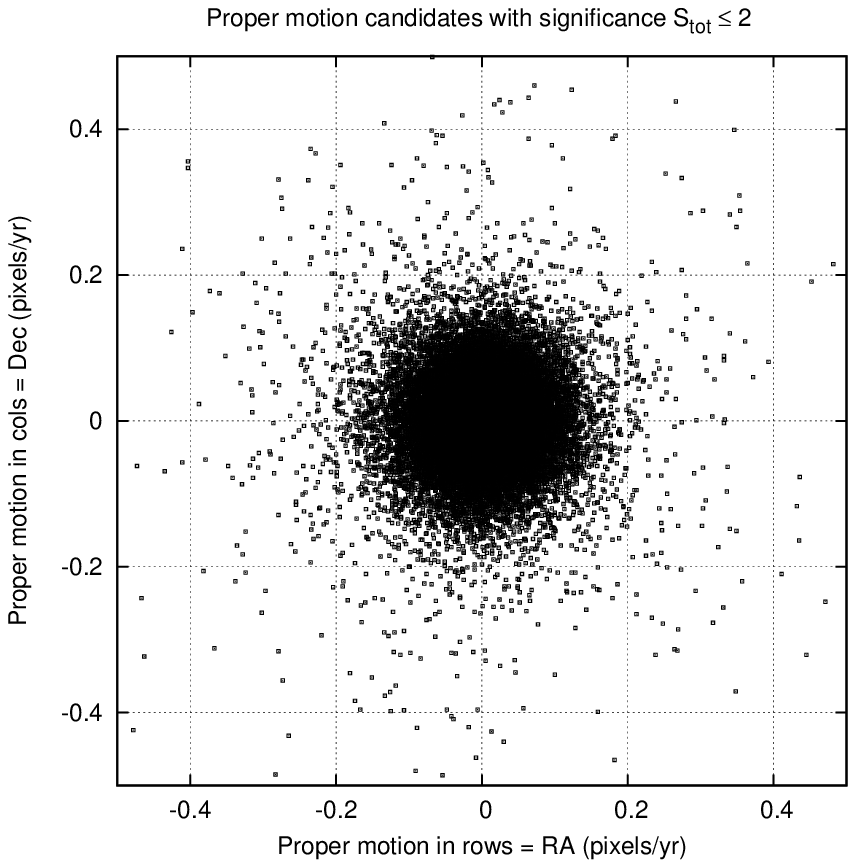}
    \FigureFile(110mm,110mm){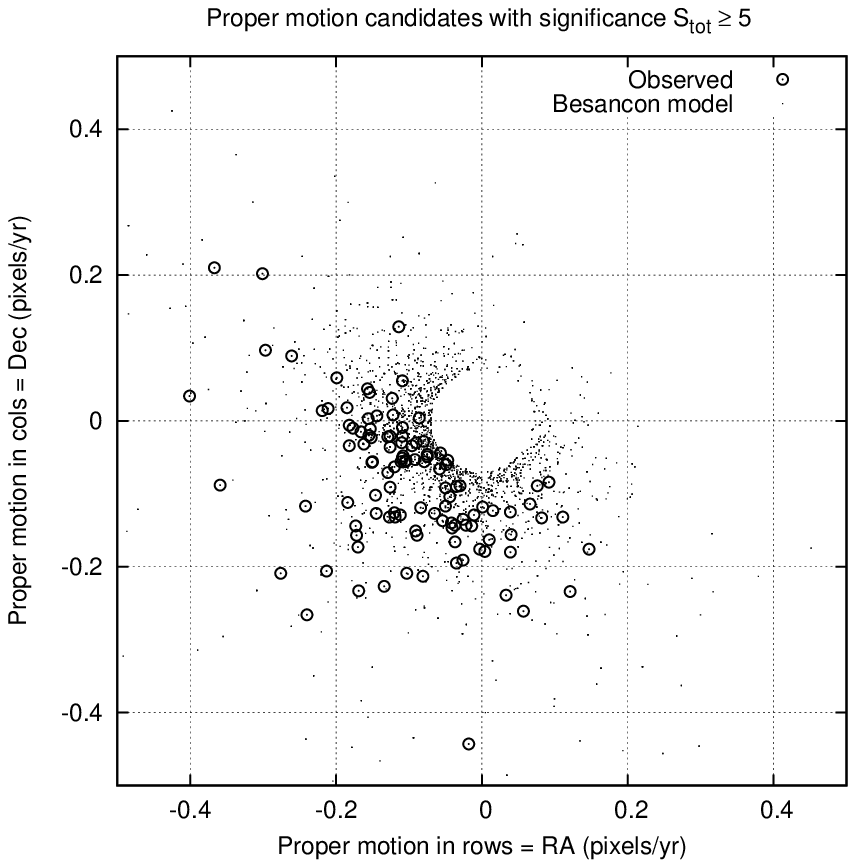}
  \end{center}
  \caption{Proper motions of objects with
             motions of small significance (top panel)
             and large significance (large circles in the bottom panel).
             \label{fig:significance} }
\end{figure}

We expect that real proper motions should show 
an asymmetry, due to the relative motions
around the galactic center
of the Sun and the stars in the SDF,
while spurious motions due to random errors in 
position measurements should be the same in all
directions.
In the upper panel of Figure \ref{fig:significance},
we plot the observed motions of objects
with motions of low significance;
they are distributed around zero with 
circular symmetry.
On the other hand,
objects with highly significant motions
(shown in the lower panel of Figure \ref{fig:significance})
are biased towards the south-east.
%
%
%
The observed asymmetry
matches that of
the stars
in 
a simulation made with
the 
Besan\c{c}on model
(the motions of which we have 
scaled appropriately).

\begin{figure}
  \begin{center}
    \FigureFile(80mm,80mm){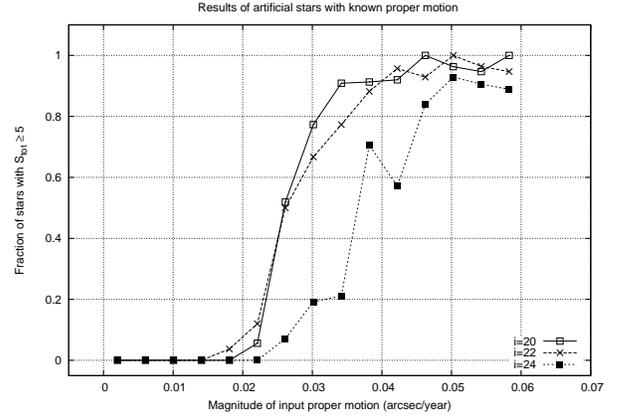}
  \end{center}
  \caption{Completeness test of the proper-motion sample
             using artificial stars.
             \label{fig:moving_signif} }
\end{figure}

For the 
sample discussed below,
we selected objects with
$S_{tot} \geq 5.0$.
Note that since our definition of 
$S$ is based on a 95-percent confidence
interval, corresponding to two standard deviations
for a normal distribution,
our criterion could be described
as ``motion at the 10-sigma level.''
Selecting objects based on the 
$S_{tot}$ statistic introduces a bias
against objects with small proper motions.
We investigated the nature of this bias
by adding artificial stars with a range of
proper motions into our images,
analyzing the images as before,
and comparing the output properties
of the artificial stars to their input values.
In 
Figure \ref{fig:moving_signif},
we show the fraction of artificial stars which 
had measured motions of high enough significance
to be included in our proper motion sample.
For bright stars, the fraction drops to 50\% at
a total proper motion of about 
$\mu = 0{\rlap.}^{''}025$ per year.

\begin{figure}
  \begin{center}
    \FigureFile(80mm,80mm){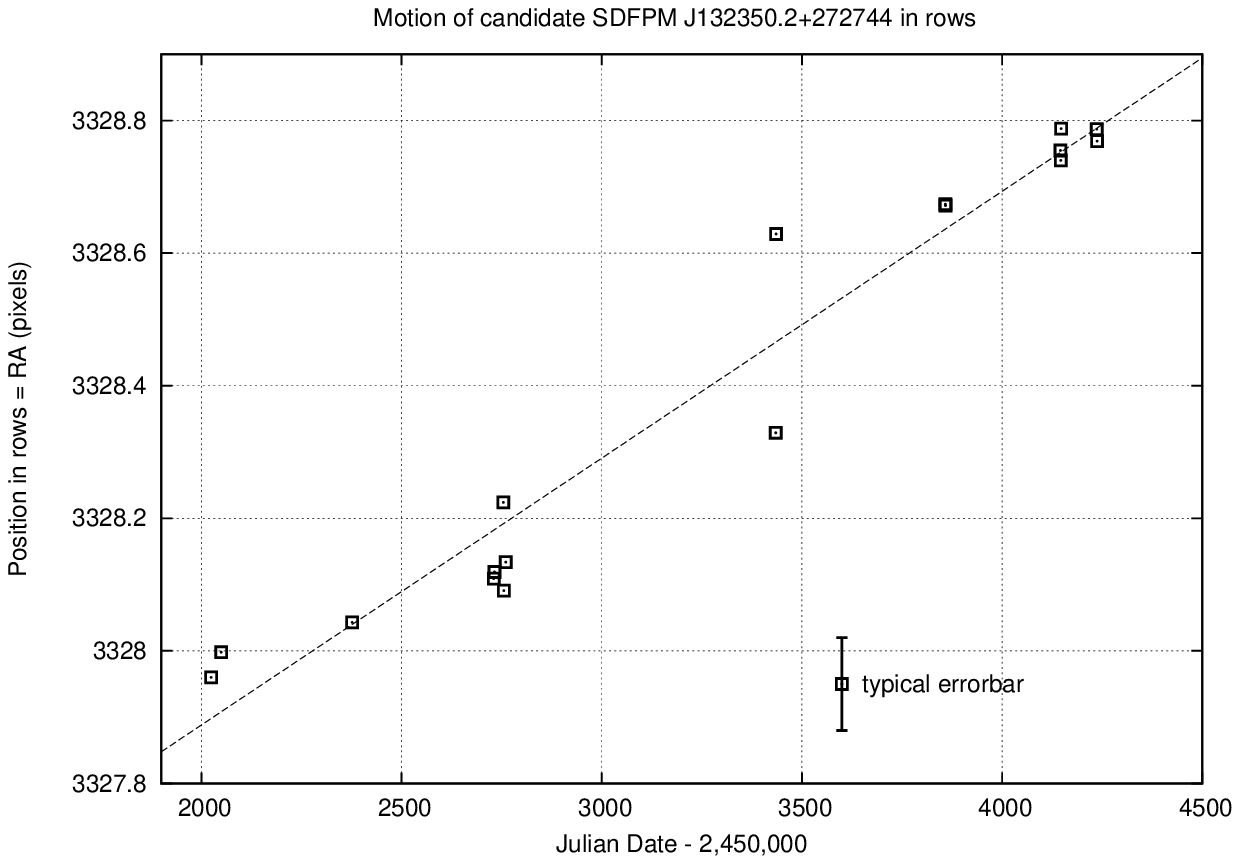}
    \FigureFile(80mm,80mm){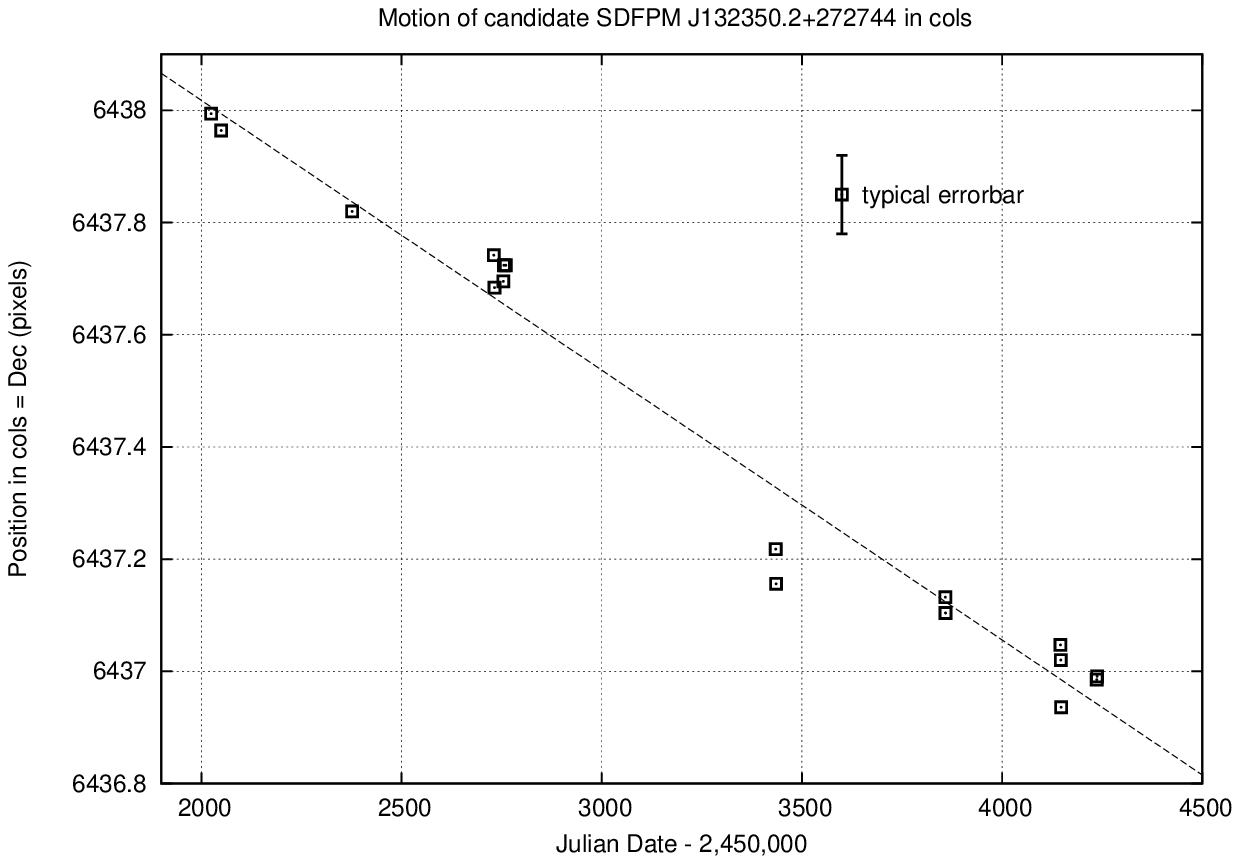}
  \end{center}
  \caption{An example of the motions for one of the
             candidates with high significance.
             \label{fig:fitted_motion} }
\end{figure}

We found that 110 objects passed this test.
However, upon visually inspecting each candidate,
we discovered that in five cases, the motions 
were due to a blend of 
two nearby stars, or a star mixed with the
light of a background galaxy.
That left a set of 105 candidates with 
real motions at a high significance.
The median number of epochs of measurement
for these objects was 19, and only 2 stars had fewer
than 15 epochs.
We show an example of the motions for one
of these candidates in 
Figure \ref{fig:fitted_motion}.

\begin{figure}
  \begin{center}
    \FigureFile(80mm,80mm){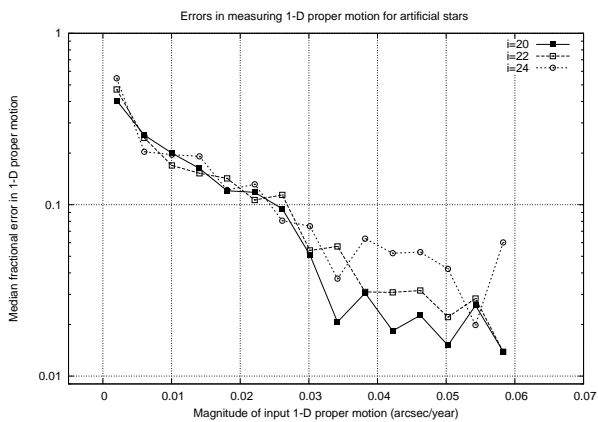}
  \end{center}
  \caption{Fractional errors in derived one-dimensional proper motions
             using artificial stars.
             \label{fig:fractional_errors} }
\end{figure}

How accurate are the derived motions?
Using our set of artificial stars again,
we computed a fractional error $E$
based on the one-dimensional motion 
of stars in row and column directions separately.
\begin{equation}
E \equiv { { ({\rm measured\ }\mu) - ({\rm input\ }\mu) } 
           \over {|{\rm input\ }\mu| + Q } }
\end{equation}
We included a constant 
$Q = 0{\rlap.}^{''}001$ per year
to prevent division by zero.
Figure \ref{fig:fractional_errors}
shows the 
median value of $E$ as a function 
of input proper motion 
for stars of different magnitudes.
The fractional error reaches 10\% 
for proper motions
of about
$\mu = 0{\rlap.}^{''}025$ per year.

Let us turn back to the
real stars in the SDF.
Figure \ref{fig:pm_histogram}
shows that the distribution of proper motions
among stars in our sample
has a peak at 
$\mu = 0{\rlap.}^{''}025$ per year,
which is
(not coincidentally)
the point at which
the efficiency
of detecting motion falls to 50\%.
The largest motions we found are about
$0{\rlap.}^{''}09$ per year,
which is far less than the limit of 
$0{\rlap.}^{''}17$ per year
set by our matching procedure.
We conclude that our matching requirement --
that each measurement lie within 
$1{\rlap.}^{''}0$ 
of its match in the fiducial epoch --
does not have a strong effect on the 
resulting proper motions.

Some of the analysis described below requires 
the color of an object, in order to distinguish
different types of star.
We chose
the $(V-I)$ color
for several reasons:
it samples a wide range of wavelengths,
avoids $B$-band measurements which are
hard to make for cool stars,
is often used in studies of 
galactic structure,
and is commonly tabulated in models
of stellar properties.
Since we used $i'$-band images to search
for motions, all the candidates had good
$i'$-band magnitudes; we will convert them to 
the standard Cousins $I$ system in 
the next paragraph.
However, many of the candidates grow faint
in images taken at shorter wavelengths.
We inspected each candidate in 
deep SDF images 
(\cite{Kashikawa2004})
taken in
$B$, $V$ and $R_C$ passbands,
and compared its appearance in those images
to the 
$B$, $V$ and $R_C$ magnitudes 
listed for each object in the SDF catalogs.
In six cases, the $V$-band measurement
was clearly incorrect,
sometimes due to confusion with a brighter 
object nearby;
the number of improper
magnitudes was much larger in $B$-band.
Removing these six objects from our sample,
we are left with a set of 99 stars
which have well measured proper motions
and good magnitudes in $V$, $R_C$, $i'$ and $z'$.
We call this our ``first sample.''
We list these candidates in
Table
\ref{table:sample}.

\begin{longtable}{l l l r r r r r r }
  \caption{First sample of proper motion candidates in the SDF}
  \label{table:sample}
  \hline \hline
  ID & RA\footnotemark[$*$] & Dec\footnotemark[$*$] & 
        $V$\footnotemark[$**$] $\thinspace $ & 
        $R_c$\footnotemark[$**$] $\thinspace $ & 
        $i'$\footnotemark[$**$] $\thinspace $ & 
        $z'$\footnotemark[$**$] $\thinspace $ & 
        RA PM\footnotemark[$\dagger$] \qquad &  
        Dec PM\footnotemark[$\dagger$] \qquad  \\ 
  \endfirsthead
  \hline \hline
  \endhead
  \hline 
  \endfoot
  \hline 
  \endlastfoot

  \hline

  SDFPM J132339.4+271916 &    200.91424 &     27.32130 & 21.784 & 21.025 & 20.164 & 19.607 &  -0.049 $\pm$  0.004 & -0.024 $\pm$  0.004 \\
  SDFPM J132343.5+272230 &    200.93138 &     27.37519 & 20.889 & 20.198 & 19.501 & 18.853 &  -0.034 $\pm$  0.004 & -0.003 $\pm$  0.006 \\
  SDFPM J132343.6+272753 &    200.93170 &     27.46494 & 22.821 & 21.995 & 20.868 & 20.265 &  -0.017 $\pm$  0.002 &  0.001 $\pm$  0.004 \\
  SDFPM J132346.9+274556 &    200.94582 &     27.76573 & 21.810 & 20.993 & 20.003 & 19.425 &  -0.073 $\pm$  0.004 & -0.018 $\pm$  0.004 \\
  SDFPM J132347.5+271829 &    200.94827 &     27.30829 & 21.711 & 21.005 & 20.505 & 20.139 &  -0.037 $\pm$  0.004 & -0.023 $\pm$  0.004 \\
  SDFPM J132348.0+273232 &    200.95033 &     27.54234 & 24.208 & 23.350 & 22.209 & 21.615 &  -0.053 $\pm$  0.002 &  0.018 $\pm$  0.002 \\
  SDFPM J132348.0+273702 &    200.95029 &     27.61729 & 23.585 & 22.739 & 21.520 & 20.867 &  -0.022 $\pm$  0.002 & -0.002 $\pm$  0.002 \\
  SDFPM J132350.2+272744 &    200.95944 &     27.46232 & 19.051 & 19.110 & 19.193 & 19.141 &   0.030 $\pm$  0.004 & -0.036 $\pm$  0.004 \\
  SDFPM J132351.0+273453 &    200.96275 &     27.58148 & 21.617 & 20.713 & 19.596 & 18.739 &   0.017 $\pm$  0.002 & -0.027 $\pm$  0.004 \\
  SDFPM J132353.1+272759 &    200.97157 &     27.46647 & 23.648 & 22.870 & 22.073 & 21.626 &  -0.006 $\pm$  0.002 & -0.018 $\pm$  0.002 \\
  SDFPM J132353.4+272207 &    200.97253 &     27.36882 & 20.417 & 19.870 & 19.564 & 19.130 &  -0.034 $\pm$  0.004 & -0.035 $\pm$  0.002 \\
  SDFPM J132354.0+272806 &    200.97518 &     27.46843 & 24.444 & 23.488 & 21.818 & 20.910 &  -0.012 $\pm$  0.000 & -0.009 $\pm$  0.002 \\
  SDFPM J132354.3+273016 &    200.97647 &     27.50454 & 23.857 & 23.334 & 22.733 & 22.322 &  -0.032 $\pm$  0.004 &  0.001 $\pm$  0.004 \\
  SDFPM J132354.5+273356 &    200.97722 &     27.56561 & 22.793 & 22.390 & 22.199 & 22.029 &  -0.056 $\pm$  0.004 & -0.042 $\pm$  0.002 \\
  SDFPM J132356.7+274445 &    200.98637 &     27.74593 & 22.248 & 21.560 & 21.150 & 20.835 &  -0.024 $\pm$  0.004 & -0.025 $\pm$  0.002 \\
  SDFPM J132357.5+271458 &    200.98977 &     27.24948 & 23.025 & 22.144 & 20.812 & 20.046 &  -0.019 $\pm$  0.002 & -0.007 $\pm$  0.002 \\
  SDFPM J132400.0+273750 &    201.00007 &     27.63083 & 21.997 & 21.169 & 20.115 & 19.443 &  -0.009 $\pm$  0.000 & -0.021 $\pm$  0.004 \\
  SDFPM J132401.1+273613 &    201.00494 &     27.60384 & 23.932 & 23.034 & 21.688 & 20.989 &  -0.036 $\pm$  0.002 & -0.002 $\pm$  0.002 \\
  SDFPM J132402.5+272639 &    201.01074 &     27.44421 & 25.483 & 24.796 & 23.022 & 22.008 &   0.003 $\pm$  0.002 & -0.025 $\pm$  0.002 \\
  SDFPM J132403.6+271314 &    201.01540 &     27.22081 & 24.830 & 24.027 & 23.312 & 22.843 &  -0.024 $\pm$  0.004 & -0.027 $\pm$  0.006 \\
  SDFPM J132403.6+273833 &    201.01540 &     27.64265 & 22.152 & 21.247 & 19.998 & 19.174 &  -0.022 $\pm$  0.004 & -0.011 $\pm$  0.002 \\
  SDFPM J132404.5+272557 &    201.01911 &     27.43275 & 24.593 & 23.464 & 21.534 & 20.563 &  -0.033 $\pm$  0.004 & -0.006 $\pm$  0.002 \\
  SDFPM J132404.5+273829 &    201.01897 &     27.64141 & 23.774 & 23.016 & 22.452 & 22.084 &  -0.023 $\pm$  0.002 & -0.026 $\pm$  0.002 \\
  SDFPM J132407.4+272751 &    201.03110 &     27.46440 & 23.326 & 22.556 & 22.030 & 21.693 &  -0.032 $\pm$  0.002 &  0.009 $\pm$  0.002 \\
  SDFPM J132407.4+272924 &    201.03107 &     27.49026 & 20.825 & 20.089 & 19.268 & 18.488 &   0.008 $\pm$  0.002 & -0.036 $\pm$  0.004 \\
  SDFPM J132407.8+273621 &    201.03275 &     27.60605 & 20.992 & 20.211 & 19.334 & 18.498 &  -0.011 $\pm$  0.002 & -0.028 $\pm$  0.002 \\
  SDFPM J132408.0+274455 &    201.03344 &     27.74886 & 23.205 & 23.035 & 23.011 & 22.952 &  -0.031 $\pm$  0.004 & -0.005 $\pm$  0.002 \\
  SDFPM J132409.7+273406 &    201.04050 &     27.56846 & 22.629 & 21.914 & 21.501 & 21.195 &  -0.018 $\pm$  0.002 & -0.032 $\pm$  0.002 \\
  SDFPM J132411.1+271310 &    201.04634 &     27.21970 & 21.478 & 20.735 & 19.868 & 19.261 &   0.001 $\pm$  0.002 & -0.036 $\pm$  0.004 \\
  SDFPM J132412.6+272220 &    201.05261 &     27.37244 & 21.385 & 20.713 & 20.279 & 19.940 &  -0.005 $\pm$  0.002 & -0.039 $\pm$  0.002 \\
  SDFPM J132413.0+272651 &    201.05433 &     27.44771 & 21.211 & 20.568 & 20.195 & 19.890 &   0.007 $\pm$  0.002 & -0.048 $\pm$  0.002 \\
  SDFPM J132413.2+273139 &    201.05511 &     27.52759 & 24.476 & 23.966 & 23.694 & 23.519 &   0.024 $\pm$  0.004 & -0.047 $\pm$  0.002 \\
  SDFPM J132413.5+271547 &    201.05657 &     27.26317 & 23.270 & 22.586 & 22.068 & 21.668 &  -0.010 $\pm$  0.002 & -0.024 $\pm$  0.002 \\
  SDFPM J132415.4+271328 &    201.06428 &     27.22462 & 23.341 & 22.562 & 21.837 & 21.368 &  -0.024 $\pm$  0.004 & -0.013 $\pm$  0.002 \\
  SDFPM J132415.9+271624 &    201.06654 &     27.27336 & 21.285 & 21.492 & 21.654 & 21.798 &  -0.037 $\pm$  0.002 &  0.004 $\pm$  0.004 \\
  SDFPM J132418.0+271902 &    201.07534 &     27.31740 & 20.754 & 20.099 & 19.576 & 19.011 &  -0.029 $\pm$  0.004 &  0.001 $\pm$  0.004 \\
  SDFPM J132419.4+273411 &    201.08115 &     27.56974 & 23.404 & 22.691 & 22.296 & 21.994 &  -0.013 $\pm$  0.002 & -0.026 $\pm$  0.004 \\
  SDFPM J132423.6+274031 &    201.09851 &     27.67550 & 24.878 & 24.092 & 22.810 & 22.094 &  -0.008 $\pm$  0.002 & -0.030 $\pm$  0.002 \\
  SDFPM J132423.7+274425 &    201.09904 &     27.74048 & 24.684 & 23.560 & 21.358 & 20.143 &  -0.023 $\pm$  0.002 &  0.026 $\pm$  0.004 \\
  SDFPM J132425.8+272415 &    201.10777 &     27.40424 & 22.134 & 21.348 & 20.242 & 19.579 &   0.015 $\pm$  0.004 & -0.018 $\pm$  0.002 \\
  SDFPM J132427.4+271919 &    201.11456 &     27.32220 & 21.667 & 20.920 & 19.939 & 19.277 &  -0.016 $\pm$  0.002 & -0.006 $\pm$  0.002 \\
  SDFPM J132427.5+274302 &    201.11464 &     27.71747 & 21.899 & 21.050 & 19.979 & 19.326 &  -0.004 $\pm$  0.000 & -0.029 $\pm$  0.004 \\
  SDFPM J132429.2+273817 &    201.12177 &     27.63832 & 22.417 & 21.510 & 20.291 & 19.581 &  -0.031 $\pm$  0.002 &  0.008 $\pm$  0.002 \\
  SDFPM J132429.7+273932 &    201.12378 &     27.65908 & 23.801 & 23.352 & 23.149 & 22.980 &  -0.019 $\pm$  0.002 & -0.011 $\pm$  0.002 \\
  SDFPM J132429.8+274304 &    201.12447 &     27.71788 & 24.576 & 23.709 & 22.521 & 21.894 &  -0.002 $\pm$  0.002 & -0.026 $\pm$  0.002 \\
  SDFPM J132430.6+272406 &    201.12789 &     27.40185 & 25.066 & 24.034 & 22.769 & 22.151 &  -0.026 $\pm$  0.002 & -0.014 $\pm$  0.004 \\
  SDFPM J132430.9+273624 &    201.12897 &     27.60693 & 23.844 & 23.145 & 22.437 & 22.015 &  -0.001 $\pm$  0.002 & -0.036 $\pm$  0.002 \\
  SDFPM J132431.3+271528 &    201.13047 &     27.25786 & 23.296 & 22.392 & 21.111 & 20.377 &  -0.008 $\pm$  0.002 & -0.028 $\pm$  0.002 \\
  SDFPM J132431.9+272236 &    201.13305 &     27.37668 & 20.942 & 20.325 & 19.957 & 19.596 &  -0.007 $\pm$  0.002 & -0.039 $\pm$  0.004 \\
  SDFPM J132432.0+273510 &    201.13343 &     27.58623 & 26.614 & 25.586 & 23.635 & 22.704 &  -0.005 $\pm$  0.002 & -0.027 $\pm$  0.002 \\
  SDFPM J132432.9+274301 &    201.13713 &     27.71701 & 21.809 & 21.082 & 20.626 & 20.313 &   0.012 $\pm$  0.004 & -0.053 $\pm$  0.002 \\
  SDFPM J132434.5+271432 &    201.14385 &     27.24230 & 21.333 & 20.533 & 19.396 & 18.506 &  -0.074 $\pm$  0.006 &  0.042 $\pm$  0.006 \\
  SDFPM J132435.0+271638 &    201.14597 &     27.27749 & 24.078 & 23.314 & 22.761 & 22.360 &  -0.022 $\pm$  0.002 & -0.010 $\pm$  0.002 \\
  SDFPM J132436.5+272345 &    201.15221 &     27.39585 & 25.624 & 24.331 & 22.079 & 20.895 &   0.022 $\pm$  0.004 & -0.027 $\pm$  0.002 \\
  SDFPM J132438.0+273622 &    201.15867 &     27.60634 & 20.594 & 20.088 & 19.836 & 19.525 &  -0.048 $\pm$  0.002 & -0.054 $\pm$  0.002 \\
  SDFPM J132438.4+273433 &    201.16001 &     27.57608 & 25.234 & 24.507 & 23.642 & 23.123 &  -0.037 $\pm$  0.004 & -0.007 $\pm$  0.006 \\
  SDFPM J132438.8+272847 &    201.16168 &     27.47980 & 21.242 & 20.387 & 19.401 & 18.580 &   0.019 $\pm$  0.004 & -0.017 $\pm$  0.002 \\
  SDFPM J132439.0+272413 &    201.16284 &     27.40377 & 21.748 & 20.978 & 20.063 & 19.473 &  -0.029 $\pm$  0.004 & -0.021 $\pm$  0.004 \\
  SDFPM J132439.2+273949 &    201.16367 &     27.66381 & 25.890 & 25.991 & 23.708 & 22.493 &  -0.025 $\pm$  0.002 &  0.006 $\pm$  0.002 \\
  SDFPM J132440.4+272911 &    201.16870 &     27.48653 & 20.569 & 20.101 & 19.872 & 19.567 &  -0.010 $\pm$  0.002 & -0.012 $\pm$  0.002 \\
  SDFPM J132440.6+272548 &    201.16954 &     27.43018 & 25.902 & 25.050 & 23.210 & 22.293 &   0.002 $\pm$  0.002 & -0.033 $\pm$  0.004 \\
  SDFPM J132440.7+271501 &    201.16992 &     27.25041 & 20.436 & 20.017 & 19.781 & 19.510 &   0.000 $\pm$  0.002 & -0.024 $\pm$  0.002 \\
  SDFPM J132444.4+273945 &    201.18514 &     27.66254 & 23.530 & 22.957 & 22.607 & 22.279 &  -0.016 $\pm$  0.002 & -0.011 $\pm$  0.002 \\
  SDFPM J132444.9+272709 &    201.18741 &     27.45264 & 22.431 & 21.622 & 20.613 & 20.051 &   0.013 $\pm$  0.004 & -0.023 $\pm$  0.002 \\
  SDFPM J132446.0+272605 &    201.19208 &     27.43499 & 22.250 & 21.553 & 21.112 & 20.813 &  -0.026 $\pm$  0.002 & -0.004 $\pm$  0.002 \\
  SDFPM J132446.8+274114 &    201.19519 &     27.68733 & 24.778 & 23.590 & 21.544 & 20.448 &  -0.040 $\pm$  0.002 &  0.012 $\pm$  0.004 \\
  SDFPM J132447.0+272814 &    201.19610 &     27.47077 & 23.045 & 22.143 & 20.783 & 20.064 &   0.008 $\pm$  0.000 & -0.025 $\pm$  0.002 \\
  SDFPM J132447.8+272157 &    201.19930 &     27.36586 & 23.236 & 22.290 & 20.836 & 19.993 &  -0.025 $\pm$  0.002 &  0.002 $\pm$  0.002 \\
  SDFPM J132448.1+272803 &    201.20046 &     27.46761 & 21.790 & 21.173 & 20.812 & 20.554 &  -0.025 $\pm$  0.002 & -0.004 $\pm$  0.002 \\
  SDFPM J132448.6+272747 &    201.20255 &     27.46314 & 25.725 & 24.967 & 23.319 & 22.436 &  -0.022 $\pm$  0.002 & -0.010 $\pm$  0.002 \\
  SDFPM J132448.8+273205 &    201.20361 &     27.53479 & 21.771 & 20.993 & 20.036 & 19.427 &  -0.081 $\pm$  0.004 &  0.007 $\pm$  0.002 \\
  SDFPM J132449.7+274510 &    201.20714 &     27.75295 & 25.447 & 24.324 & 22.278 & 21.207 &  -0.043 $\pm$  0.002 & -0.042 $\pm$  0.002 \\
  SDFPM J132451.7+273110 &    201.21556 &     27.51972 & 25.980 & 25.510 & 25.241 & 24.948 &  -0.021 $\pm$  0.008 & -0.042 $\pm$  0.004 \\
  SDFPM J132452.1+271813 &    201.21743 &     27.30386 & 24.103 & 23.165 & 21.596 & 20.792 &  -0.025 $\pm$  0.002 & -0.007 $\pm$  0.004 \\
  SDFPM J132453.1+271821 &    201.22154 &     27.30592 & 22.338 & 21.642 & 21.242 & 20.944 &  -0.035 $\pm$  0.002 & -0.029 $\pm$  0.004 \\
  SDFPM J132454.0+274226 &    201.22508 &     27.70739 & 23.462 & 22.682 & 21.698 & 21.148 &  -0.012 $\pm$  0.002 & -0.013 $\pm$  0.002 \\
  SDFPM J132455.3+272957 &    201.23056 &     27.49939 & 25.095 & 24.307 & 23.423 & 22.968 &  -0.022 $\pm$  0.002 & -0.004 $\pm$  0.004 \\
  SDFPM J132456.0+274126 &    201.23345 &     27.69078 & 25.805 & 24.801 & 22.578 & 21.065 &  -0.060 $\pm$  0.002 &  0.020 $\pm$  0.002 \\
  SDFPM J132456.1+272807 &    201.23412 &     27.46872 & 22.942 & 22.059 & 20.673 & 19.936 &  -0.030 $\pm$  0.004 & -0.011 $\pm$  0.000 \\
  SDFPM J132458.1+272326 &    201.24241 &     27.39072 & 23.633 & 22.700 & 21.372 & 20.676 &  -0.022 $\pm$  0.002 &  0.011 $\pm$  0.002 \\
  SDFPM J132459.8+271251 &    201.24932 &     27.21436 & 23.548 & 22.849 & 22.374 & 21.961 &  -0.022 $\pm$  0.002 & -0.006 $\pm$  0.006 \\
  SDFPM J132500.3+272357 &    201.25147 &     27.39927 & 22.401 & 21.675 & 21.122 & 20.775 &  -0.015 $\pm$  0.002 & -0.010 $\pm$  0.002 \\
  SDFPM J132503.8+273938 &    201.26620 &     27.66072 & 21.497 & 20.741 & 19.999 & 19.479 &  -0.018 $\pm$  0.002 & -0.006 $\pm$  0.002 \\
  SDFPM J132504.6+273028 &    201.26917 &     27.50779 & 21.614 & 20.808 & 19.792 & 19.074 &  -0.021 $\pm$  0.002 & -0.011 $\pm$  0.000 \\
  SDFPM J132505.4+273731 &    201.27253 &     27.62530 & 22.684 & 21.943 & 21.160 & 20.728 &  -0.017 $\pm$  0.002 & -0.024 $\pm$  0.002 \\
  SDFPM J132505.3+271440 &    201.27212 &     27.24446 & 22.854 & 22.143 & 21.642 & 21.233 &  -0.007 $\pm$  0.002 & -0.029 $\pm$  0.002 \\
  SDFPM J132505.5+273401 &    201.27318 &     27.56696 & 22.398 & 21.772 & 21.336 & 21.058 &  -0.021 $\pm$  0.002 & -0.011 $\pm$  0.002 \\
  SDFPM J132506.1+273816 &    201.27567 &     27.63800 & 21.999 & 21.309 & 20.880 & 20.594 &  -0.007 $\pm$  0.002 & -0.034 $\pm$  0.002 \\
  SDFPM J132508.4+273553 &    201.28501 &     27.59830 & 20.672 & 20.018 & 19.546 & 19.011 &  -0.003 $\pm$  0.002 & -0.029 $\pm$  0.004 \\
  SDFPM J132512.6+271620 &    201.30271 &     27.27244 & 22.398 & 21.731 & 21.362 & 21.034 &  -0.009 $\pm$  0.000 & -0.011 $\pm$  0.000 \\
  SDFPM J132512.9+274045 &    201.30410 &     27.67940 & 25.443 & 24.610 & 23.500 & 22.863 &  -0.015 $\pm$  0.002 & -0.009 $\pm$  0.002 \\
  SDFPM J132514.3+272421 &    201.30981 &     27.40588 & 21.422 & 20.499 & 19.379 & 18.486 &  -0.007 $\pm$  0.002 & -0.018 $\pm$  0.002 \\
  SDFPM J132514.7+272642 &    201.31146 &     27.44522 & 23.121 & 22.559 & 22.187 & 21.881 &  -0.025 $\pm$  0.004 & -0.018 $\pm$  0.006 \\
  SDFPM J132514.7+271707 &    201.31146 &     27.28543 & 22.993 & 22.180 & 21.211 & 20.640 &  -0.027 $\pm$  0.002 & -0.046 $\pm$  0.004 \\
  SDFPM J132515.3+274212 &    201.31400 &     27.70335 & 22.304 & 22.297 & 22.357 & 22.406 &  -0.004 $\pm$  0.004 & -0.089 $\pm$  0.002 \\
  SDFPM J132515.7+272708 &    201.31560 &     27.45226 & 24.696 & 23.629 & 21.645 & 20.570 &  -0.031 $\pm$  0.002 & -0.012 $\pm$  0.002 \\
  SDFPM J132516.9+274518 &    201.32065 &     27.75507 & 23.660 & 22.886 & 22.281 & 21.891 &   0.008 $\pm$  0.004 & -0.032 $\pm$  0.004 \\
  SDFPM J132521.1+271927 &    201.33814 &     27.32435 & 25.853 & 25.090 & 24.331 & 23.887 &  -0.018 $\pm$  0.004 & -0.031 $\pm$  0.006 \\
  SDFPM J132525.4+273755 &    201.35591 &     27.63220 & 22.179 & 21.330 & 20.105 & 19.355 &  -0.031 $\pm$  0.002 & -0.004 $\pm$  0.002 \\
  SDFPM J132527.7+274407 &    201.36559 &     27.73528 & 25.738 & 24.934 & 23.663 & 22.972 &  -0.016 $\pm$  0.002 & -0.043 $\pm$  0.004 \\
  SDFPM J132527.7+272350 &    201.36576 &     27.39737 & 22.596 & 21.898 & 21.498 & 21.218 &  -0.010 $\pm$  0.002 & -0.019 $\pm$  0.002 \\
  SDFPM J132528.3+274355 &    201.36796 &     27.73214 & 25.676 & 24.699 & 22.785 & 21.825 &  -0.043 $\pm$  0.002 &  0.003 $\pm$  0.002 \\
  SDFPM J132528.3+272012 &    201.36811 &     27.33679 & 22.163 & 21.424 & 20.568 & 20.040 &  -0.031 $\pm$  0.004 & -0.002 $\pm$  0.002 \\
  SDFPM J132533.6+274708 &    201.39036 &     27.78563 & 23.894 & 22.665 & 20.660 & 19.533 &  -0.061 $\pm$  0.008 &  0.041 $\pm$  0.006 \\
  SDFPM J132533.9+272808 &    201.39152 &     27.46905 & 23.653 & 22.925 & 22.398 & 22.030 &  -0.029 $\pm$  0.002 & -0.026 $\pm$  0.002 \\

\hline
\multicolumn{8}{@{}l@{}}{\hbox to 0pt{\parbox{180mm}{\footnotesize
\par\noindent
\footnotemark[$*$] Equinox J2000, epoch 2007.13.
\par\noindent
\footnotemark[$**$] Corrected isophotal magnitudes from catalogs of
                     \citet{Kashikawa2004}.
\par\noindent
\footnotemark[$\dagger$] Proper motions in arcseconds per year.
}\hss}}

\end{longtable}

Before we can compare our measurements to models
of galactic structure, we need to convert
the Suprime-Cam $i'$ magnitudes,
which are calibrated on the AB system 
(
\cite{Fukugita1996};
\cite{Miyazaki2002};
\cite{Kashikawa2004}),
to $I$ magnitudes,
which are on the standard Johnson-Cousins system.
We used synthetic photometry to find the 
relationship between 
the Suprime-Cam $(V_s - i')$
and Johnson-Cousins
$(V-I)$,
taking bandpasses from 
\citet{Miyazaki2002}
and
\citet{Bessell1990},
respectively.
We selected main sequence stars,
O5V to M6V, from
the library of 
\citet{Pickles1998},
stars ranging in metallicity 
$ -2 \leq [{\rm Fe/H}] \leq 0$ 
from models of
\citet{Lejeune1997},
and flux-calibrated spectra of white dwarfs
observed by the SDSS
(\cite{Adelman-McCarthy2008}).
We convolved each spectrum with the
Suprime-Cam passbands and with the Johnson-Cousins
passbands 
to compute synthetic magnitudes,
and used the spectrum of Vega
from 
\citet{Bohlin2004}
to set the zeropoints to the values given in
\citet{Fukugita1996}.
We found that the following linear relationship fit
the data well, yielding a scatter of less
than 0.03 mag across the range of colors
$-0.3 < (V - I) < 3.5$:
\begin{equation}
(V-I) = 0.391 + 1.1145(V_s - i')
\end{equation}
We use this equation to convert
the observed colors for stars in the
SDF to Johnson-Cousins $(V-I)$ 
when comparing our results to stellar models.

\section{Simple analysis of the first sample}

We begin with the 
Besan\c{c}on
model.
We generated 10 simulated catalogs of objects
in the area of the SDF, using the 
parameters found by
\citet{Robin2003}
and including stars down to an apparent
magnitude of $V = 30$.
We applied cuts to the synthetic catalogs
to match the combined limits of the $i'$-band proper motion images
and the SDF catalogs.

\begin{eqnarray}
V &\leq  26.0 \\
I &\leq  25.4 \\
0{\rlap.}^{''}014 &\leq \mu \leq 0{\rlap.}^{''}17 
\end{eqnarray}

The result should be a set of stars similar
to those in the actual SDF,
though ten times more numerous.
The large size of this synthetic sample
will make it easier to delineate 
sparsely populated regions in
the reduced proper motion diagram,
to which we now turn.

Reduced proper motion was introduced by 
\citet{Luyten1922}
as a way to separate stars of different luminosities
using only the observable 
apparent magnitude, $m$,
and proper motion, $\mu$, in units of arcseconds per year.
We will base our reduced proper motion
on $V$-band magnitudes, so that

\begin{equation}
H_V = m_V + 5 \log(\mu) + 5
\end{equation}

It is also possible to express this quantity
in terms of a star's absolute magnitude, $M$,
and tangential velocity, $v_t$, expressed in units of km/s,

\begin{equation}
H_V = M_V + 5 \log(v_t) - 3.378
\end{equation}

\begin{figure*}
  \begin{center}
    \FigureFile(150mm,240mm){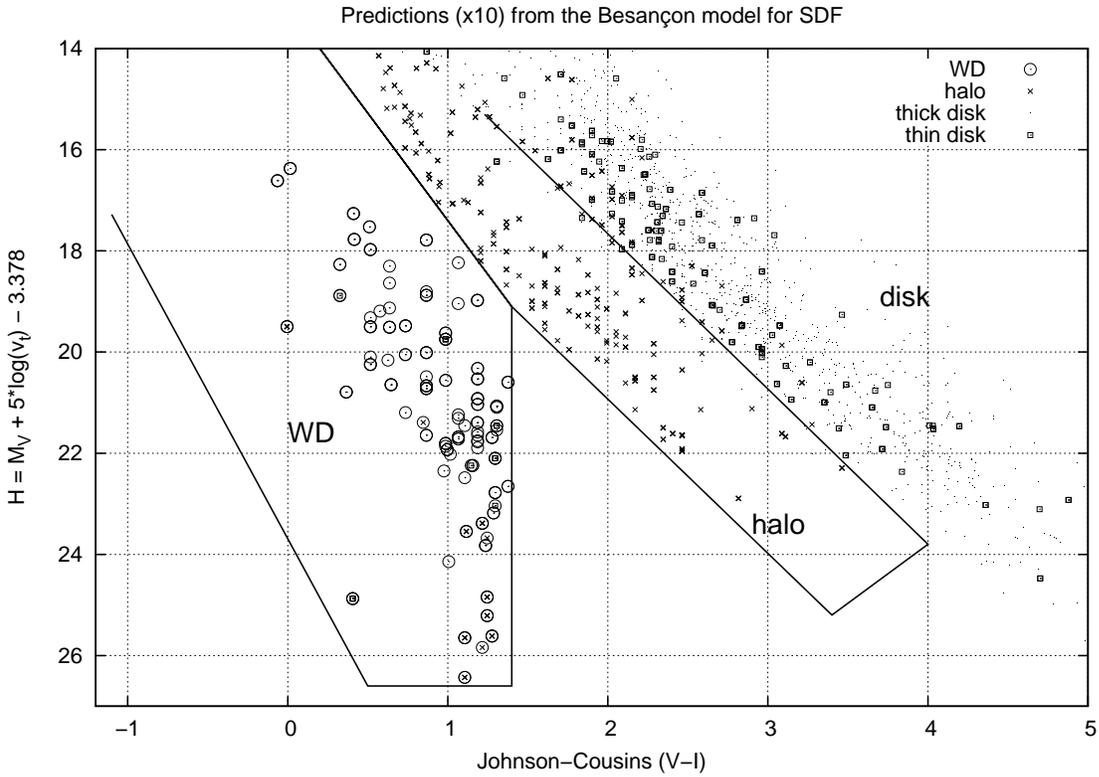}
  \end{center}
  \caption{Reduced proper motion diagram for objects in 
             simulated catalogs created with the 
             Besan\c{c}on 
             model \label{fig:reduced_besancon} }
\end{figure*}

Using this version of the formula,
we compute $H_V$ for stars
in the simulated catalogs produced from
the Besan\c{c}on model.
Figure \ref{fig:reduced_besancon}
shows the reduced proper motion as a function
of $(V-I)$ color.
We assigned objects in the 
simulation to three populations based on 
their metallicity and the component of their
space velocity in the direction of galactic rotation,
which we denote as $vgr$ to avoid confusion with
the passband $V$.

\begin{itemize}
\item{ if $vgr < -130 {\rm \ km/s}$ and  $[Fe/H] < -1.20$,
            we assign the star to the {\bf halo } }
\item{ if $vgr > -60 {\rm \ km/s}$ and  $[Fe/H] > -0.50$,
            we assign the star to the {\bf thin disk } }
\item{ otherwise, we assign the star to the {\bf thick disk } }
\end{itemize}
Stars from different populations appear in distinct
regions in this diagram.
We have drawn rough outlines by hand to aid the reader
in recognizing the populations.

Note that there is a clear ``red edge'' in
the distribution of WDs, at a color of
$(V-I) \sim 1.4$.
Theoretical models of cooling WDs
(\cite{Richer2000};
\cite{Chabrier2000})
indicate that at an age of about
10 Gyr and a temperature of
$T_{eff} \sim 5000 {\rm \ K}$,
an increase in opacity due to molecular
hydrogen causes the $(V-I)$ color 
to shift back to the blue
as the star continues to cool.
WDs with atmospheres dominated by
other elements, such as helium or carbon,
would continue to grow redder 
as they cool.
Objects near 
the bottom 
of the WD region 
are likely to be members of the halo.

\begin{figure*}
  \begin{center}
    \FigureFile(150mm,240mm){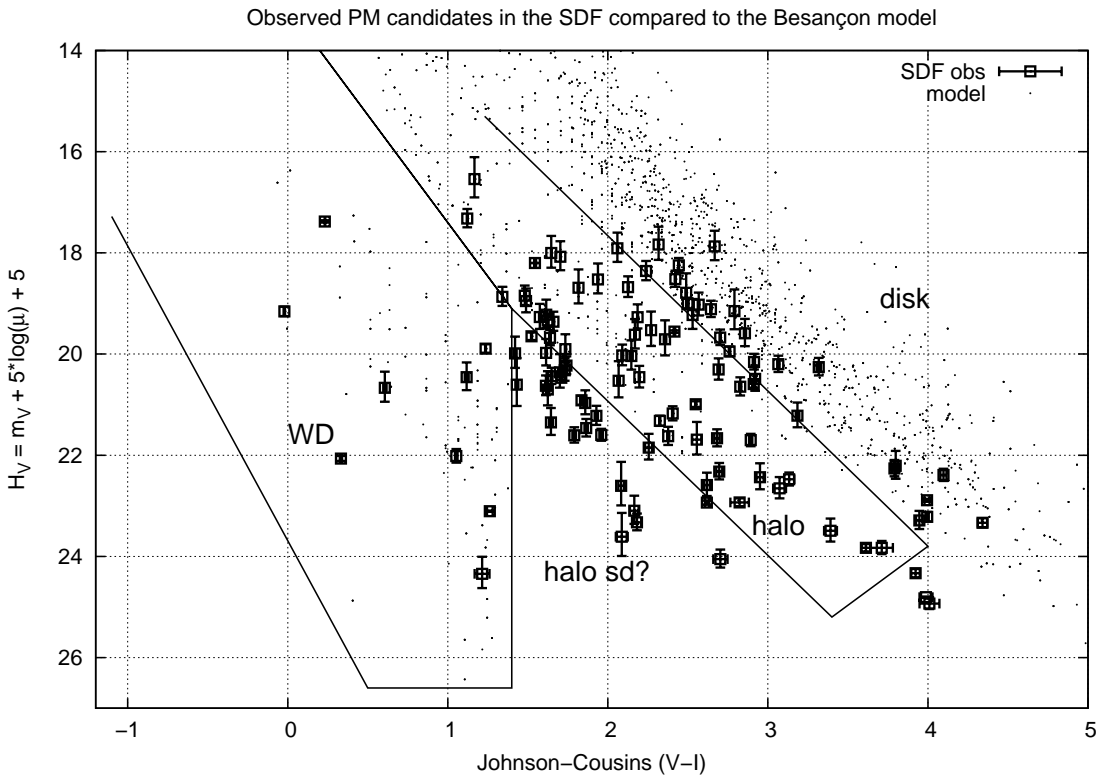}
  \end{center}
  \caption{Reduced proper motion diagram for 
             real objects observed in the SDF.
             The regions are identical to those in
             Figure \ref{fig:reduced_besancon}.
             Each measurement has errorbars in both directions,
             though some are too small to see.
             \label{fig:reduced_sdf} }
\end{figure*}

In Figure \ref{fig:reduced_sdf},
we present the reduced proper motions
for the real stars in our first sample.
We must switch to the first form of reduced
proper motion,
Equation 7,
to compute $H_V$ for the observed stars.
To facilitate comparison with the
Besan\c{c}on 
model,
we include the hand-drawn regions
from 
Figure \ref{fig:reduced_besancon}
as well as all objects from the 
simulated catalogs as tiny points.
Note that the saturation of very bright stars
$V \lesssim 20$
in the Subaru images,
plus our limited ability to measure proper motions
$\mu \lesssim 0{\rlap.}^{''}02$ 
per year,
combine to eliminate
any candidates
with reduced proper motions $H \lesssim 16.5$.
In the discussion which follows, 
please recall that the boundaries 
of the regions drawn in the diagram 
are only approximations intended
to provide rough classifications;
the number of items within each
region could change by ten or twenty
percent if one shifted the boundaries
slightly.

Our 99 proper motion (PM) candidates
divide into four groups:
9 fall inside
the WD region, 43 inside the 
halo region, 
23 in the disk region,
and 24 lie in
an area which had no stars in the 
Besan\c{c}on 
model.
Let us discuss the WD candidates first,
and then consider the objects in the
``empty'' area.

The PM candidates falling into the WD region
are concentrated near the red edge of
the region, just as models of WD cooling
predict.
We take the combined results for
ten simulated catalogs generated from the
Besan\c{c}on 
model,
correct for completeness 
as a function of magnitude and proper motion,
based on our tests with artificial stars
(see 
Figure \ref{fig:moving_signif}),
and divide by 10 to find predictions
of $2.2$ WDs in the halo, $5.9$ in the thick disk
and $0.7$ in the thin disk, for a total of $8.8$
WDs satisfying our selection criteria in the SDF.
Our sample yields 9 candidates in this
region, consistent with the model.
Note that several candidates lie just outside
the WD region;
we must make additional measurements of these
objects before we can make any confident claim
about the exact number of WDs.
Which of our candidates are most likely to be
members of the halo?
As a young WD cools, it slides diagonally down
and to the right on this diagram, parallel to
the lower envelope of the simulation's objects.
Due to their high velocities, halo WDs should lie
near the bottom of the distribution.
There are several candidates at some distance below the main 
locus of simulation objects; we believe 
the two at 
$H_V = 23.0$ and 
$H_V = 24.2$
are the most likely of our candidates to be halo dwars.

Roughly one-quarter of our PM candidates
lie in an ``empty'' region between the
simulation's WD and halo stars.
Since we drew the boundaries by hand,
they may certainly be shifted by small
amounts;
that would cause a number of the 
anamolous objects to fall within 
the halo or WD regions.
However, some of our candidates
are more than one magnitude away from 
any population in the simulations.
Is it possible that they may be ordinary stars
reddened by dust?
According to 
\citet{Schlegel1998},
dust along the line of sight through the 
SDF should produce $E(B-V) = 0.015$ mag;
the corresponding extinction, $A_V = 0.051$ and $A_I = 0.030$, 
is too small to shift candidates
a significant distance in the reduced PM diagram.
We suggest that these objects
may be metal-poor subdwarfs in the halo,
a class of star which is not included
in the 
Besan\c{c}on 
model.
If we make an HR diagram using
$M_V$ and $(V-I)$, 
we find that the halo stars in the 
Besan\c{c}on 
model cross the disk main sequence
in the range $1.5 < (V-I) < 3.0$,
and are slightly more luminous in 
the redder portion of this range.
However, as 
\citet{Reid1998} show,
extremely metal-poor subdwarfs
are much fainter than disk stars in this color range,
by up to 4 magnitudes.
Stars with these photometric properties
and halo kinematics
would lie several magnitudes below the
halo region drawn in our diagrams.
We therefore tentatively identify 
as extremely metal-poor 
halo stars
the PM candidates which fall
far from the WD and halo regions.

\section{Future work}

Our first step will be to acquire spectra of some of our
PM candidates to verify their identity as WDs and
metal-poor subdwarfs.
After we have spectra for a good fraction of our 
candidates, we can assign types to the candidates
with more confidence.
At that point, we will make a more detailed
and quantitative comparison of the WDs in our
sample with those expected from models of the Milky Way.

We can also look at candidates with motions of
somewhat lower significance.
In this paper, we examined the 110 stars which
had $S_{tot} > 5.0$.
There were 72 stars
with $4.0 \leq S_{tot} < 5.0$,
and they show nearly the same degree of
asymmetry in their motions 
as our first sample.
It is likely that a significant number of these stars
have real proper motions.
However, it will
take extra effort to distinguish them from the growing
number of false detections,
and to check their photometry (many of them are fainter than
the stars in our sample).
As we 
sift through this set of stars with
less significant motions,
we can also try to improve the $B$ and $V$
measurements of our stars,
so that we will not discard so many candidates
due to their uncertain colors.

Finally, we can apply the same techniques to find
stars with large proper motions in
other deep fields with multiple visits by Subaru.
Our next target will be the area of the
Subaru/XMM-Newton Deep Survey 
(\cite{Sekiguchi2004},
\cite{Furusawa2008}), which is roughly
in the opposite direction from the SDF.
We can use it to check our results in the SDF,
but also to look for differences predicted by
models of galactic populations.

%
%

We thank the staff at the Subaru Telescope 
for their assistance with the observations used in this project.
MWR gratefully acknowledges grant S-03031 from the
JSPS Invitation Fellowship for Research in Japan.
This work is supported in part by a JSPS core-to-core program
``International Research Network for Dark Energy'' and by
a JSPS research grant (17104002).
Long ago, Charles Alcock suggested that we use Suprime-Cam 
to search for white dwarfs; we are glad to be putting his
idea into practice at last.
A.G. acknowledges support by the Benoziyo Center for Astrophysics
and the William Z. and Eda Bess Novick New Scientists Fund at the
Weizmann Institute. 

Funding for the Sloan Digital Sky Survey (SDSS) and SDSS-II has been provided
by the Alfred P. Sloan Foundation, the Participating Institutions, the National
Science Foundation, the U.S. Department of Energy, the National Aeronautics and
Space Administration, the Japanese Monbukagakusho, and the Max Planck Society,
and the Higher Education Funding Council for England. The SDSS Web site is
http://www.sdss.org/.

The SDSS is managed by the Astrophysical Research Consortium (ARC) for the
Participating Institutions. The Participating Institutions are the American
Museum of Natural History, Astrophysical Institute Potsdam, University of
Basel, University of Cambridge, Case Western Reserve University, The University
of Chicago, Drexel University, Fermilab, the Institute for Advanced Study, the
Japan Participation Group, The Johns Hopkins University, the Joint Institute
for Nuclear Astrophysics, the Kavli Institute for Particle Astrophysics and
Cosmology, the Korean Scientist Group, the Chinese Academy of Sciences
(LAMOST), Los Alamos National Laboratory, the Max-Planck-Institute for
Astronomy (MPIA), the Max-Planck-Institute for Astrophysics (MPA), New Mexico
State University, Ohio State University, University of Pittsburgh, University
of Portsmouth, Princeton University, the United States Naval Observatory, and
the University of Washington.

\end{document}